\PassOptionsToPackage{unicode}{hyperref}
\PassOptionsToPackage{hyphens}{url}
\documentclass[11pt,british,english]{article}
\usepackage[T1]{fontenc}
\usepackage{geometry}
\usepackage{xcolor}
\usepackage{babel}
\usepackage{amsmath}
\usepackage{amsthm}
\usepackage{amssymb}
\usepackage{setspace}
\usepackage[authoryear]{natbib}
\PassOptionsToPackage{normalem}{ulem}
\usepackage{ulem}
\usepackage[unicode=true,
bookmarks=true,bookmarksnumbered=true,bookmarksopen=false,
breaklinks=true,pdfborder={0 0 1},backref=false,colorlinks=true]
{hyperref}
\hypersetup{pdftitle={Identifiability and Estimation of Structural Singular Vector Autoregressive Models},
	pdfauthor={Funovits, Braumann}}

\makeatletter

\providecolor{lyxadded}{rgb}{0,0,1}
\providecolor{lyxdeleted}{rgb}{1,0,0}

\DeclareRobustCommand{\lyxsout}[1]{\ifx\\#1\else\sout{#1}\fi}

\theoremstyle{plain}
\newtheorem{lem}{\protect\lemmaname}
\theoremstyle{remark}
\newtheorem{rem}{\protect\remarkname}
\theoremstyle{plain}
\newtheorem{prop}{\protect\propositionname}

\date{}

\makeatother

\addto\captionsbritish{\renewcommand{\lemmaname}{Lemma}}
\addto\captionsbritish{\renewcommand{\propositionname}{Proposition}}
\addto\captionsbritish{\renewcommand{\remarkname}{Remark}}
\addto\captionsenglish{\renewcommand{\lemmaname}{Lemma}}
\addto\captionsenglish{\renewcommand{\propositionname}{Proposition}}
\addto\captionsenglish{\renewcommand{\remarkname}{Remark}}
\providecommand{\lemmaname}{Lemma}
\providecommand{\propositionname}{Proposition}
\providecommand{\remarkname}{Remark}

\usepackage{lmodern}
\usepackage{amsmath}
\usepackage{ifxetex,ifluatex}
\ifnum 0\ifxetex 1\fi\ifluatex 1\fi=0 %
  \usepackage[T1]{fontenc}
  \usepackage[utf8]{inputenc}
  \usepackage{textcomp} %
  \usepackage{amssymb}
\else %
  \usepackage{unicode-math}
  \defaultfontfeatures{Scale=MatchLowercase}
  \defaultfontfeatures[\rmfamily]{Ligatures=TeX,Scale=1}
\fi
\IfFileExists{upquote.sty}{\usepackage{upquote}}{}
\IfFileExists{microtype.sty}{%
  \usepackage[]{microtype}
  \UseMicrotypeSet[protrusion]{basicmath} %
}{}
\makeatletter
\@ifundefined{KOMAClassName}{%
  \IfFileExists{parskip.sty}{%
    \usepackage{parskip}
  }{%
    \setlength{\parindent}{0pt}
    \setlength{\parskip}{6pt plus 2pt minus 1pt}}
}{%
  \KOMAoptions{parskip=half}}
\makeatother
\usepackage{xcolor}
\IfFileExists{xurl.sty}{\usepackage{xurl}}{} %
\IfFileExists{bookmark.sty}{\usepackage{bookmark}}{\usepackage{hyperref}}
\hypersetup{
  pdftitle={Supplement to `Identifiability of Structural Singular Vector Autoregressive Models'},
  pdfauthor={Bernd Funovits, Alexander Braumann},
  hidelinks,
  pdfcreator={LaTeX via pandoc}}
\usepackage{color}
\usepackage{fancyvrb}

\DefineVerbatimEnvironment{Highlighting}{Verbatim}{commandchars=\\\{\}}
\usepackage{framed}
\definecolor{shadecolor}{RGB}{248,248,248}
\newenvironment{Shaded}{\begin{snugshade}}{\end{snugshade}}

\newcommand{\AttributeTok}[1]{\textcolor[rgb]{0.77,0.63,0.00}{#1}}

\newcommand{\CommentTok}[1]{\textcolor[rgb]{0.56,0.35,0.01}{\textit{#1}}}

\newcommand{\ConstantTok}[1]{\textcolor[rgb]{0.00,0.00,0.00}{#1}}
\newcommand{\ControlFlowTok}[1]{\textcolor[rgb]{0.13,0.29,0.53}{\textbf{#1}}}

\newcommand{\DecValTok}[1]{\textcolor[rgb]{0.00,0.00,0.81}{#1}}
\newcommand{\DocumentationTok}[1]{\textcolor[rgb]{0.56,0.35,0.01}{\textbf{\textit{#1}}}}

\newcommand{\FloatTok}[1]{\textcolor[rgb]{0.00,0.00,0.81}{#1}}
\newcommand{\FunctionTok}[1]{\textcolor[rgb]{0.00,0.00,0.00}{#1}}
\newcommand{\ImportTok}[1]{#1}

\newcommand{\NormalTok}[1]{#1}
\newcommand{\OperatorTok}[1]{\textcolor[rgb]{0.81,0.36,0.00}{\textbf{#1}}}
\newcommand{\OtherTok}[1]{\textcolor[rgb]{0.56,0.35,0.01}{#1}}

\newcommand{\SpecialCharTok}[1]{\textcolor[rgb]{0.00,0.00,0.00}{#1}}

\newcommand{\StringTok}[1]{\textcolor[rgb]{0.31,0.60,0.02}{#1}}
\newcommand{\VariableTok}[1]{\textcolor[rgb]{0.00,0.00,0.00}{#1}}

\usepackage{graphicx}
\makeatletter
\def\maxwidth{\ifdim\Gin@nat@width>\linewidth\linewidth\else\Gin@nat@width\fi}
\def\maxheight{\ifdim\Gin@nat@height>\textheight\textheight\else\Gin@nat@height\fi}
\makeatother
\setkeys{Gin}{width=\maxwidth,height=\maxheight,keepaspectratio}
\makeatletter
\def\fps@figure{htbp}
\makeatother
\providecommand{\tightlist}{%
  \setlength{\itemsep}{0pt}\setlength{\parskip}{0pt}}
\ifluatex
  \usepackage{selnolig}  %
\fi

\title{Supplement to `Identifiability of Structural Singular Vector
Autoregressive Models'}
\author{Bernd Funovits, Alexander Braumann}
\date{}

\begin{document}
	\newgeometry{left=4cm,right=4cm,top=0cm,bottom=0.2cm}
	\thispagestyle{empty}
	\title{Identifiability of Structural Singular Vector Autoregressive Models}
	\author{Bernd Funovits$^{a,b}$ and Alexander Braumann$^{c}$}
	
	\maketitle
	\thispagestyle{empty}
	\vspace{-2cm}
	
	\section*{}
	
	\section*{Affiliations}
	
	\begin{singlespace}
		\textbf{$^{a}$University of Helsinki}\\
		Faculty of Social Sciences\\
		Discipline of Economics\\
		P. O. Box 17 (Arkadiankatu 7)\\
		FIN-00014 University of Helsinki\\
	\end{singlespace}
	and
	\begin{singlespace}
		\textbf{$^{b}$TU Wien}\\
		Institute of Statistics and Mathematical Methods in Economics\\
		Econometrics and System Theory\\
		Wiedner Hauptstr. 8\\
		A-1040 Vienna
	\end{singlespace}
	
	and
	
	\begin{singlespace}
			\textbf{$^{c}$TU Braunschweig}\\
		Institute for Mathematical Stochastics\\
		Universitätsplatz 2\\
		D-38106 Braunschweig\\
	\end{singlespace}
	
	\section*{E-mail of Corresponding Author}
	
	bernd.funovits@helsinki.fi
	
	\pagebreak{}
	
	\thispagestyle{empty}
	\setcounter{page}{0}
	\newgeometry{left=3cm,right=3cm,top=3cm,bottom=3cm}
	
	\section*{Abstract}
	
	We generalize well-known results on structural identifiability of
	vector autoregressive (VAR) models to the case where the innovation
	covariance matrix has reduced rank. Singular structural VAR models
	appear, for example, as solutions of rational expectation models where
	the number of shocks is usually smaller than the number of endogenous
	variables, and as an essential building block in dynamic factor models.
	We show that order conditions for identifiability are misleading in
	the singular case and we provide a rank condition for identifiability
	of the noise parameters. Since the Yule-Walker (YW) equations may
	have multiple solutions, we analyze the effect of restrictions on
	the system parameters on over- and underidentification in detail and
	provide easily verifiable conditions.
	
	\textbf{Keywords:} Stochastic singularity, structural vector autoregressive
	models, identifiability
	
	\textbf{JEL classification: }C32, C50
	
	\pagebreak{}
	
	\section{Introduction}
	
	Singular structural VAR (SVAR) models play an important role in macroeconomic
	modeling. To introduce the subject, we succinctly discuss Generalized
	Dynamic Factor Models (GDFM) and Dynamic Stochastic General Equilibrium
	(DSGE) models, and their relation to singular SVAR models.
	
	In the literature on GDFMs \citep{FHLR00,FHLR05,BaiNg07,EJC10}, singular
	VAR models are the essential building block connecting static factors
	(a static transformation of the denoised observables) to the uncorrelated
	lower-dimensional shocks. \citet{chenandersondeistlerfiller2010}
	and \citet{festschriftbdoanderson2010} treat canonical forms of singular
	VAR models, i.e. they focus on the reduced form. In \citet{FGLR05},
	it is demonstrated that dynamic factor models (and consequently singular
	VAR models) are useful for structural modeling. In this article, we
	provide results regarding identifiability of singular SVAR models
	and thus analyze Step C in \citet[page 1332]{FGLR05} in more detail.
	
	A key issue in the econometric treatment of DSGE models is caused
	by the fact that the number of exogenous shocks driving the system
	is often strictly smaller than the number of endogenous variables.
	This is known as the stochastic singularity problem \citep[page 184f.]{DeJongDave11}
	and investigated in, e.g., \citet{RugeMurcia07singularDSGE}. The
	relationship between DSGE and SVAR models is analyzed in \citet{DeJongDave11},
	\citet{Giacomini13}, \citet[Chapter 6.2]{KilianLut17}, and most
	recently by \citet{lippi19}. It has been acknowledged \citep[page 177]{KilianLut17}
	that the usual strategies\footnote{\citet{KilianLut17} enumerate 1) adding measurement noise as, e.g.
		in \citet{Sargent89} or \citet{Ireland04}, and discussed in \citet{lippi19},
		2) reducing the number of observables \citep{bouakezRugemurcia2005habit}
		and 3) augmenting the number of economically interpretable shocks
		\citep{IngramKocherlotkaSaven94,LeeperSims94}.} for solving this rank deficiency problem are not satisfactory. Thus,
	one way forward would be estimation of singular SVAR models.
	
	Singularity of the innovation covariance matrix has two possible consequences
	for the restrictions imposed by the modeler\footnote{A similar problem appears in \citet[Chapter 10.2]{KilianLut17} where
		it is emphasized that the reduced rank of a certain matrix appearing
		in cointegration analysis must be ``taken into account when determining
		the number of restrictions that have to be imposed for full identification
		of the structural shocks''.}. On the one hand, the restrictions imposed by the modeler might contradict
	the restrictions that are implicit due to the singularity structure
	of the innovation covariance matrix. On the other hand, the restrictions
	imposed by the modeler might already be contained in the restrictions
	that are implicit due to the singularity structure of the innovation
	covariance matrix and are therefore redundant. These cases must be
	taken into account when analyzing identifiability properties of singular
	SVAR models. Moreover, restrictions on the system parameters are not
	necessarily over-identifying when the innovation covariance matrix
	is singular because the YW equations might have multiple solutions.
	
	The rest of this article is structured as follows. In section 2, we
	specify the model, we introduce restrictions on model parameters in
	a general fashion, and we define notions which will be necessary later.
	As preparation for the main results, we collect well-known facts on
	singular VAR models in reduced form (in particular regarding the possible
	non-singularity of the Toeplitz matrix appearing in the YW equations)
	in section 3. In section 4, we analyze restrictions on the noise and
	system parameters and provide results as to how a singular innovation
	covariance matrix needs to be taken into account for identifiability
	analysis. We illustrate that the usual order condition may be misleading
	in the singular case with a (stochastically singular) DSGE model and
	provide easily verifiable conditions for under- and over-identification
	when the YW equations have multiple solutions.\textcolor{red}{{} }All
	proofs are deferred to the Appendix.
	
	The following notation is used in the article. We use $z$ as a complex
	variable as well as the backward shift operator on a stochastic process,
	i.e. $z\left(y_{t}\right)_{t\in\mathbb{Z}}=\left(y_{t-1}\right)_{t\in\mathbb{Z}}$.
	For a (matrix) polynomial $p(z)$, we denote by $\deg(p(z))$ the
	highest degree of $p(z)$. The transpose of an $\left(m\times n\right)$
	dimensional matrix $A$ is represented by $A'$. We use ${\rm vec}\left(A\right)\in\mathbb{R}^{nm\times1}$
	to stack the columns of $A$ into a column vector and ${\rm vech}\left(A\right)\in\mathbb{R}^{\frac{n(n+1)}{2}\times1}$
	to stack the lower-triangular elements of an $n$-dimensional square
	matrix $A$ analogously. The $n$-dimensional identity matrix is denoted
	by $I_{n}$. The inequality $">0"$ refers to positive definiteness
	in the context of matrices. For the span of the row space and the
	column space of $A$, we write $span_{R}\left(A\right)$ and $span_{C}\left(A\right)$,
	respectively, and the projection of $A$ on $span_{R}\left(B\right)$,
	$B\in\mathbb{R}^{r\times n}$, is $Proj_{R}\left(A|B\right)$ and
	the projection of $A$ on $span_{C}\left(D\right)$, $D\in\mathbb{R}^{m\times s}$,
	is $Proj_{C}\left(A|D\right)$. We use $\mathbb{E}\left(\cdot\right)$
	for the expectation of a random variable with respect to a given probability
	space.
	
	\section{Model}
	
	Here, we start by defining the model, i.e. the system and noise parameters
	as well as the stability, singularity, and researcher imposed restrictions.
	Next, we describe the observed quantities that are available to the
	econometrician; in our case the second moments. Last, we discuss the
	notion of identifiability, i.e. the connection between the internal
	and external characteristics.
	
	We consider a SVAR\footnote{Most work on SVAR models is performed in the parametrization where
		$A_{0}=I_{n}$ \citep[Chapter 8]{KilianLut17} and investigates how
		to estimate $B$. Since we also treat structural restrictions on
		$A_{+}$, rather than on the reduced form parameters $A_{0}^{-1}A_{+}$,
		we allow here for additional generality and treat the so-called $AB$-model
		(in the nomenclature of \citet{luet05}). Except for section 4.2,
		it is sufficient to set $A_{0}=I_{n}$. } system
	\begin{align}
		A_{0}y_{t} & =A_{1}y_{t-1}+\cdots+A_{p}y_{t-p}+B\varepsilon_{t},\label{eq:model}\\
		& =\underbrace{\left(A_{1},\ldots,A_{p}\right)}_{=A_{+}}x_{t-1}+B\varepsilon_{t}\nonumber 
	\end{align}
	where $x_{t-1}=\left(y'_{t-1},\ldots,y'_{t-p}\right)'$ and where
	the dimension $q$ of the white noise process $(\varepsilon_{t})$
	of (economically) fundamental shocks with covariance matrix $I_{q}$
	is strictly smaller than $n$, the number of observed variables of
	$y_{t}$. The matrix $B\in\mathbb{R}^{n\times q}$ has full column
	rank. This implies that the covariance matrix $\Sigma_{u}$ of the
	innovations $u_{t}=B\varepsilon_{t}$ is of rank $q<n$.
	
	Furthermore, we assume that the matrices $A_{i}\in\mathbb{R}^{n\times n}$
	are such that the stability condition 
	\begin{equation}
		\det\left(a(z)\right)\neq0,\ \left|z\right|\leq1,\label{eq:restr_stability}
	\end{equation}
	holds, where $a(z)=A_{0}-A_{1}z-\cdots-A_{p}z^{p}$, and that $\det\left(A_{0}\right)\neq0$.
	Last, we assume that the system and noise parameters satisfy the restrictions
	\begin{equation}
		C_{S}{\rm vec}\left(A_{+}'\right)=c_{S}\text{ and }C_{N}{\rm vec}\left(\begin{pmatrix}A_{0} & B\end{pmatrix}\right)=c_{N}\label{eq:restr_modeler}
	\end{equation}
	where $C_{S}$ and $C_{N}$ are of dimensions $\left(r_{S}\times n^{2}p\right)$
	and $\left(r_{N}\times\left(n^{2}+nq\right)\right)$, respectively,
	describing the (a-priori known) restrictions imposed by the modeler.
	To summarize, we define the internal characteristics that we would
	like to identify as the parameters $\left(A_{+},\left(A_{0},B\right)\right)$
	in system (\ref{eq:model}) which satisfy the restrictions imposed
	by (\ref{eq:restr_stability}) and (\ref{eq:restr_modeler}).
	
	Next, we discuss the external characteristics which are observed by
	the econometrician. The stationary solution of the system (\ref{eq:model})
	(together with the restrictions imposed on the parameters) is called
	a singular VAR process. Having available all finite joint distributions
	of the singular VAR process corresponds to the maximal information
	we could possibly obtain regarding external characteristics. Another
	commonly used set of external characteristics is the second moment
	information contained in the singular VAR process, i.e. the autocovariance
	function $\gamma(s)=\mathbb{E}\left(y_{t}y_{t-s}'\right)$ or equivalently
	the spectral density $f\left(e^{-i\lambda}\right)=\frac{1}{2\pi}\sum_{s=-\infty}^{\infty}\gamma(s)e^{-is\lambda}$.
	
	We follow \citet{Rothenberg71} to define identifiability of parametric
	models. Two internal characteristics $\left(A_{+}^{(1)},\left(A_{0}^{(1)},B^{(1)}\right)\right)$
	and $\left(A_{+}^{(2)},\left(A_{0}^{(2)},B^{(2)}\right)\right)$ are
	called observationally equivalent if they imply the same external
	characteristics. An internal characteristic is globally identifiable
	if there is no other observationally equivalent internal characteristic.
	Likewise, an internal characteristic $\left(A_{+},\left(A_{0},B\right)\right)$
	is locally identifiable if there exists a neighborhood around the
	parameter $\left(A_{+},\left(A_{0},B\right)\right)$ corresponding
	to the internal characteristic such that there is no other observationally
	equivalent internal characteristic in this neighborhood. In this article,
	we focus on identifiability from second moment information, i.e. the
	external characteristics correspond to the spectral density of the
	observed process $\left(y_{t}\right)$.

	\section{Identifiability Issues in Reduced Form Singular VAR Models \label{sec:YuleWalker}}
	
	In order to prepare for the structural case where we will connect
	to external characteristics uniquely to the deep parameters, we review
	identifiability of the reduced form of singular VAR models, see also
	\citet{andersondeistlerchenfiller2012}. In particular, we discuss
	the rank of finite sections of the covariance of the observed process
	and its relation to the rank of the innovation covariance matrix $\Sigma_{u}$.
	Moreover, we show how $p$, $q$ and the left-kernel $L\in\mathbb{R}^{(n-q)\times n}$
	of $\Sigma_{u}$, which are assumed to be known in the identifiability
	analysis in Section \ref{sec:structural_identifiability}, can be
	obtained from the external characteristics.
	
	One way to connect the observable characteristics to the internal
	characteristics is by using the YW equations\footnote{They are obtained by right-multiplying $\left(y_{t}',\ldots,y_{t-p}'\right)$
		on (\ref{eq:model}) and taking expectations.}, i.e. 
	\[
	\bar{A}_{+}\Gamma_{p}=\gamma_{p}\ \text{and}\ \Sigma_{u}=\gamma(0)-\bar{A}_{+}\gamma_{p}',
	\]
	where $\bar{A}_{+}=\left(\bar{A}_{1},\ldots,\bar{A}_{p}\right)=A_{0}^{-1}A_{+}$
	are the reduced form system parameters, $\Gamma_{p}=\left(\begin{smallmatrix}\gamma(0) & \gamma(1) & \cdots & \gamma(p-1)\\
		\gamma(-1) & \gamma(0)\\
		\vdots &  & \ddots\\
		\gamma(-p+1) &  &  & \gamma(0)
	\end{smallmatrix}\right)$ and $\gamma_{p}=\left(\gamma(1),\ldots,\gamma(p)\right)$. If $\Gamma_{p}$
	is invertible, there is a unique internal characteristic $\left(\bar{A}_{+},\Sigma_{u}\right)$
	for a given external characteristic $\left(\gamma(0),\ldots,\gamma(p)\right)$.
	While for VAR models with non-singular innovation covariance matrix
	it can be shown \citep[page 112]{HannanDeistler12} that $\Gamma_{r}$
	is non-singular for all $r\in\mathbb{N}$, this is not the case for
	VAR models that have a singular innovation covariance matrix. Indeed,
	it is easy to see \citep[Theorem 7]{andersondeistler09} that, for
	$s>0$, $rk\left(\Gamma_{p+s}\right)=rk\left(\Gamma_{p}\right)+s\cdot q$
	holds. Even $\Gamma_{p}$ might be rank deficient: Consider a solution
	$\left(\bar{A}_{+}^{(1)},\Sigma_{u}\right)$ of the YW equations and
	the polynomial matrix $U(z)=I_{n}+cc'z$ where $c'\in\mathbb{R}^{1\times n}$
	is non-trivial and in the left-kernel of both $\bar{A}_{p}^{(1)}$
	and $\bar{B}=A_{0}^{-1}B$. One can verify that $U(z)\bar{a}^{(1)}(z)$,
	where $\bar{a}^{(1)}(z)$ is the polynomial corresponding to $\bar{A}_{+}^{(1)}$,
	is also a polynomial matrix of degree $p$ and solves the YW equations
	which implies that $\Gamma_{p}$ has a non-trivial left-kernel. Note
	that the perpendicular of the projection is unique irrespective of
	how the projection itself is parametrized. More formally, $\Sigma_{u}^{(1)}=\gamma(0)-\bar{A}_{+}^{(1)}\gamma_{p}'=\gamma(0)-\bar{A}_{+}^{(2)}\gamma_{p}'=\Sigma_{u}^{(2)}$
	holds even if $\bar{A}_{+}^{(1)}\neq\bar{A}_{+}^{(2)}$ for two solutions
	$(\bar{A}_{+}^{(1)},\Sigma_{u}^{(1)})$ and $(\bar{A}_{+}^{(2)},\Sigma_{u}^{(2)})$
	of the Yule-Walker equations. 
	
	Examining the ranks of $\Gamma_{r}$ for some consecutive values of
	$r$, the integer-valued parameters $q$ and $p$ can be obtained.
	Having the rank $q$ of the innovation covariance $\Sigma_{u}$ available,
	it is straightforward to obtain (a basis of) the left-kernel $L\in\mathbb{R}^{(n-q)\times n}$
	of $\Sigma_{u}$ \citep{AlSadoon17ranktests}.
	
	For the remainder of this article, we will assume that $p,q,$ and
	$L$ are known by the practitioner (in addition to the other external
	characteristics).

	\section{\label{sec:structural_identifiability}Imposing Structural Restrictions}
	
	In this section, we discuss identifiability of noise and system parameters
	in the case of singular SVAR models. First, we derive a condition
	which ensures that the modeler imposed restrictions on the noise parameters
	are not in contradiction to the singularity of the innovation covariance
	matrix. Subsequently, we derive a rank condition similar to the previous
	literature and illustrate with a New-Keynesian DSGE model that the
	order condition does not provide useful information in the stochastically
	singular case. Secondly, we discuss whether researcher imposed restrictions
	on system parameters are under-, just- or over-identifying. In particular,
	we show that it is uncommon that researcher imposed restrictions do
	not solve the underidentification problem (if the number of restrictions
	is at least as large as the rank deficiency of $\left(I_{n}\otimes\Gamma_{p}\right)$).
	
	We start with affine restrictions on the noise parameters $\left(A_{0},B\right)$
	which appear in short-run restrictions, see \citet[Chapter 8]{KilianLut17}
	for the non-singular case. The conditions that we derive are local
	in nature. Next, we deal with the case where $\Gamma_{p}$ may be
	singular and where affine restrictions on the elements in $A_{+}$
	are imposed. These results concern global identifiability.

	\subsection{Affine Restrictions on the Noise Parameters}
	
	In the light of the discussion in Section \ref{sec:YuleWalker}, we
	start with a singular $\Sigma_{u}$ and with researcher imposed restrictions
	given by 
	\begin{equation}
		C_{N}{\rm vec}\left(\left(A_{0},B\right)\right)=c_{N}.\label{eq:restr_B_affine}
	\end{equation}
	Here, $C_{N}=\left(\begin{smallmatrix}C_{A_{0}} & 0_{r_{A_{0}}\times nq}\\
		0_{r_{B}\times n^{2}} & C_{B}
	\end{smallmatrix}\right)$ is block-diagonal and has full row rank, and $c'_{N}=\left(c'_{A_{0}},c'_{B}\right)$.
	In order to show the existence of a unique pair $(A_{0},B)$ for parametrising
	$\Sigma_{u}=A_{0}^{-1}BB'\left(A_{0}^{'}\right)^{-1}$, one usually
	calls on the implicit functions theorem. While in the non-singular
	SVAR case the system of equations to be analyzed always has at least
	one solution, it might happen in the singular SVAR case that the set
	of solutions of (\ref{eq:restr_B_affine}) (for which the restrictions
	imposed by the researcher are satisfied) is the empty set. Since the
	premises of the implicit function theorem are such that there must
	be at least one solution, one needs to make sure that the affine restrictions
	(\ref{eq:restr_B_affine}) imposed by the researcher do not contradict
	the singularity structure of the model. In the following, we will
	provide an analytical condition which implies and is implied by a
	non-empty solution set.
	
	The linear dependence structure induced by the singularity of $\Sigma_{u}$
	implies $L\left(A_{0}^{-1}B\right)=0$, where the rows of $L\in\mathbb{R}^{(n-q)\times n}$
	span the left-kernel of $\Sigma_{u}$, which is equivalent to
	\begin{equation}
		\left(I_{q}\otimes L\right){\rm vec}\left(A_{0}^{-1}B\right)=0.\label{eq:restr_B_singul}
	\end{equation}
	The condition for when the solution set of the joint system of restrictions
	given in (\ref{eq:restr_B_affine}) and (\ref{eq:restr_B_singul})
	is non-empty is given in the following 
	\begin{lem}
		\label{lem:rest_affine_compatible}Let $L\in\mathbb{R}^{(n-q)\times n}$
		be a basis of the left-kernel of $\Sigma_{u}$, define $\mathcal{N}:=\left\{ \left[\left(A_{0}^{-1}B\right)',I_{q}\right]\otimes LA_{0}^{-1}\right\} $,
		and let $M:=C_{N}-Proj_{R}(C_{N}|\mathcal{N})$ be the perpendicular
		of the projection of $C_{N}$ on the row-span of $\mathcal{N}$. The
		restrictions $C_{N}{\rm vec}\left(A_{0},B\right)=c_{N}$ are consistent
		with the singularity of $\Sigma_{u}$ if and only if ${\rm rk}(M)={\rm rk}\begin{pmatrix}M & c_{N}\end{pmatrix}$,
		i.e. if and only if $c_{N}$ is in the image of $M$.
	\end{lem}
	\begin{rem}
		When we consider the SVAR setting in which $A_{0}=I_{n}$, we only
		need to check whether $c_{B}$ is contained in the column space of
		$C_{B}-Proj\left(C_{B}|\left(I_{q}\otimes L\right)\right)$.
	\end{rem}
	The singularity of $\Sigma_{u}$ restricts the set of admissible restrictions
	on the parameter space. If $C_{N}$ does not ``interfere'' with
	the singularity restrictions, i.e. if $C_{N}$ lies in the orthogonal
	complement of $span_{R}\left(\mathcal{N}\right)$ or expressed differently
	if $Proj_{R}\left(C_{B}|\mathcal{N}\right)=0$, then $M=C_{N}$ and
	condition ${\rm rk}(M)={\rm rk}\begin{pmatrix}M & c_{N}\end{pmatrix}$
	is satisfied.
	
	\begin{prop}
		\label{prop:BmodelLocalIdent}Let $A_{0}$ and $B$ be $\left(n\times n\right)$
		and $\left(n\times q\right)$-dimensional matrices of full column
		rank, let $n>q$, and let $C_{N}{\rm vec}\left(A_{0},B\right)=c_{N}$
		hold. For given $\Sigma_{u}$, the matrix $\left(A_{0},B\right)$
		is the unique solution of $\Sigma_{u}=A_{0}^{-1}B\left(A_{0}^{-1}B\right)'$
		if and only if $c_{N}$ is in the image of $M=C_{N}-Proj_{R}\left(C_{N}|\mathcal{N}\right)$
		and the matrix $\left(\begin{smallmatrix}-2D_{n}^{+}(\Sigma_{u}\otimes A_{0}^{-1}) & 2D_{n}^{+}(A_{0}^{-1}B\otimes A_{0}^{-1})\\
			C_{A_{0}} & 0\\
			0 & C_{B}
		\end{smallmatrix}\right)$ is of (full column) rank $n^{2}+nq$.
	\end{prop}
	\begin{rem}
		Considering for simplicity the case where $A_{0}=I_{n}$ and following
		\citet{Rothenberg71}, the restrictions imposed on the structural
		parameter $B$ are $C_{B}{\rm vec}(B)=c_{B}$ as well as $\left(I_{q}\otimes L\right)vec(B)=0$
		which suggests that the matrix $\frac{\partial}{\partial\left(vec\left(B\right)\right)'}\left(\begin{smallmatrix}{\rm vec}h\left(BB'\right)-{\rm vech}\left(\Sigma_{u}\right)\\
			C_{B}{\rm vec}(B)-c_{B}\\
			\left(I_{q}\otimes L\right){\rm vec}(B)
		\end{smallmatrix}\right)$ needs to be of rank $nq$. However, it is not necessary to include
		$\left(I_{q}\otimes L\right)$ in Proposition \ref{prop:BmodelLocalIdent}
		because $\left(I_{q}\otimes L\right){\rm vec}(B)=0$ is already implied
		by the fact that $BB'=\Sigma_{u}$. Put differently, the inequality
		$rk\left(\begin{smallmatrix}2D_{n}^{+}(B\otimes I_{n})\\
			C_{B}\\
			\left(I_{q}\otimes L\right)
		\end{smallmatrix}\right)\leq rk\left(\begin{smallmatrix}2D_{n}^{+}(B\otimes I_{n})\\
			C_{B}
		\end{smallmatrix}\right)$ holds.
	\end{rem}
	\begin{rem}
		If $q<n$, the usual order condition requiring that the number of
		rows in $\left(\begin{smallmatrix}2D_{n}^{+}\left(B\otimes I_{n}\right)\\
			C_{B}
		\end{smallmatrix}\right)$ be larger than or equal to the number of columns is not useful. Consider
		the case where there are no researcher imposed restrictions. While
		the order condition is satisfied for $q\leq\frac{n+1}{2}$, the matrix
		$D_{n}^{+}\left(B\otimes I_{n}\right)$ of dimension $\left(\frac{n(n+1)}{2}\times nq\right)$
		is of course rank deficient with co-rank $\frac{q(q-1)}{2}$.
	\end{rem}
	\begin{rem}
		The rank of the matrix $\left(\begin{smallmatrix}2D_{n}^{+}\left(B\otimes I_{n}\right)\\
			C_{B}
		\end{smallmatrix}\right)$ drops if some restrictions in $C_{B}$ are already implied by the
		singularity structure of $\Sigma_{u}$, i.e. if for the $r$-th row
		$\left[C_{B}\right]_{\left[r,\bullet\right]}\subseteq{\rm span_{R}}\left(I_{q}\otimes L\right)$
		holds. Thus, the $\frac{q(q-1)}{2}$ additional restrictions which
		are necessary to obtain a matrix $\left(\begin{smallmatrix}2D_{n}^{+}\left(B\otimes I_{n}\right)\\
			C_{B}
		\end{smallmatrix}\right)$ of full column rank must not be contained in the row space of $D_{n}^{+}\left(B\otimes I_{n}\right)$.
	\end{rem}

	\subsubsection{Illustration}
	
	To illustrate Proposition \ref{prop:BmodelLocalIdent}, we discuss
	a version of the New-Keynesian monetary business cycle model \citep{LubikSchorfheide03,castelnuovo13}
	featuring a ``supply-shifting'' shock in the new-Keynesian Phillips
	Curve (NKPC). We thus consider the model 
	\begin{align*}
		\pi_{t} & =\beta\mathbb{E}_{t}\left(\pi_{t+1}\right)+\kappa x_{t}+\varepsilon_{t}^{\pi}\\
		x_{t} & =\mathbb{E}_{t}\left(x_{t+1}\right)-\tau\left(R_{t}-\mathbb{E}_{t}\left(\pi_{t+1}\right)\right)\\
		R_{t} & =\phi\mathbb{E}_{t}\left(\pi_{t+1}\right)+\varepsilon_{t}^{R}.
	\end{align*}
	where $\left(\pi_{t},x_{t},R_{t}\right)$ denote inflation, output
	gap, and nominal interest rate in log-deviation from a unique steady
	state. The conditional expectations are to be understood as linear
	projections on the space spanned by present and past components of
	the uncorrelated shocks $\varepsilon_{t}^{\pi}$ and $\varepsilon_{t}^{R}$
	which are white noise processes (whose variance is normalized to one
	for the sake of simplicity). The parameters of the model are the subjective
	time preference factor $\beta\in\left(0,1\right)$, $\phi\geq0$ the
	elasticity of the interest response of the central bank, and the slope
	parameters $\kappa$ and $\tau$.
	
	For specific parameter values $(\beta,\phi,\tau,\kappa)=\left(\frac{4}{5},\frac{39}{38},\frac{3}{4},\frac{1}{2}\right)$,
	we solve this system of equations involving conditional expectations
	of future endogenous variables \citep{Sims01,funo17full} and obtain
	the unique causal stationary solution $\left(\begin{smallmatrix}R_{t}\\
		\pi_{t}\\
		x_{t}
	\end{smallmatrix}\right)=B\left(\begin{smallmatrix}\varepsilon_{t}^{R}\\
		\varepsilon_{t}^{\pi}
	\end{smallmatrix}\right)$, where $B=\left(\begin{smallmatrix}1 & 0\\
		-\kappa\tau & 1\\
		-\tau & 0
	\end{smallmatrix}\right)$, of the DSGE model described above. The innovation covariance matrix
	is obviously singular. The restrictions on $B$ are described by
	$C_{B}{\rm vec}(B)=c_{B}$, with $C_{B}=\left(\begin{smallmatrix}1 & 0 & 0 & 0 & 0 & 0\\
		0 & 0 & 0 & 1 & 0 & 0\\
		0 & 0 & 0 & 0 & 1 & 0\\
		0 & 0 & 0 & 0 & 0 & 1
	\end{smallmatrix}\right)$ and $c_{B}=\left(\begin{smallmatrix}1\\
		0\\
		1\\
		0
	\end{smallmatrix}\right)$. In order to apply Proposition \ref{prop:BmodelLocalIdent}, we need
	to check the condition ${\rm rk}(M)={\rm rk}\begin{pmatrix}M & c_{B}\end{pmatrix}$
	of Lemma \ref{lem:rest_affine_compatible}. The perpendicular of the
	projection of $C_{B}$ on the row-span of $\left(I_{2}\otimes L\right)$
	for $L=\begin{pmatrix}\tau & 0 & 1\end{pmatrix}$, is given by $M=\left(\begin{smallmatrix}\frac{1}{1+\tau^{2}} & 0 & -\frac{\tau}{1+\tau^{2}} & 0 & 0 & 0\\
		0 & 0 & 0 & \frac{1}{1+\tau^{2}} & 0 & -\frac{\tau}{1+\tau^{2}}\\
		0 & 0 & 0 & 0 & 1 & 0\\
		0 & 0 & 0 & -\frac{\tau}{1+\tau^{2}} & 0 & \frac{\tau^{2}}{1+\tau^{2}}
	\end{smallmatrix}\right)$. For any value $\tau\in\mathbb{R}\backslash\left\{ 0\right\} $ the
	relation ${\rm rk}(M)={\rm rk}\begin{pmatrix}M & c_{B}\end{pmatrix}$
	is satisfied. We can now apply Proposition \ref{prop:BmodelLocalIdent}
	and check the rank of $\left(\begin{smallmatrix}2D_{3}^{+}(B\otimes I_{3})\\
		C_{B}
	\end{smallmatrix}\right)=\left(\begin{smallmatrix}2 & 0 & 0 & 0 & 0 & 0\\
		-\kappa\tau & 1 & 0 & 1 & 0 & 0\\
		-\tau & 0 & 1 & 0 & 0 & 0\\
		0 & -2\kappa\tau & 0 & 0 & 2 & 0\\
		0 & -\tau & -\kappa\tau & 0 & 0 & 1\\
		0 & 0 & -2\tau & 0 & 0 & 0\\
		1 & 0 & 0 & 0 & 0 & 0\\
		0 & 0 & 0 & 1 & 0 & 0\\
		0 & 0 & 0 & 0 & 1 & 0\\
		0 & 0 & 0 & 0 & 0 & 1
	\end{smallmatrix}\right)$ is equal to 6 for any values $\kappa$ and $\tau$.
	
	\subsection{Affine Restrictions on the System Parameters}
	
	We now focus on imposing linear restrictions on the structural parameters
	$A_{+}$ in the case where $\Gamma_{p}$ is singular. Thus, without
	restrictions on $A_{+}$, there are multiple observationally equivalent
	solutions of the YW equations (one particular solution plus the left
	kernel of $\Gamma_{p}$). We will start by considering the case where
	$A_{0}=I_{n}$ (such that the reduced form parameters $\bar{A}_{+}$
	coincide with $A_{+}$). This simplifies the discussion and allows
	us to illustrate why the identifiability problem (for $A_{0}$ not
	necessarily equal to the identity matrix) can ``generically'' be
	solved by (the right number of) arbitrary restrictions on $A_{+}$.\footnote{To be more precise, it can be considered uncommon that $s\cdot n$,
		where $s$ is the dimension of the kernel of $\Gamma_{p}$, ``random''
		restrictions on ${\rm vec}\left(A_{+}'\right)$ do not solve the identifiability
		problem.}
	
	Two aspects deserve special attention. First, the particular solutions
	(canonical representatives of the equivalence class of observational
	equivalence) introduced in \citet{festschriftbdoanderson2010} and
	\citet{chenandersondeistlerfiller2010} can be obtained by choosing
	a particular set of restrictions on ${\rm vec}\left(A_{+}'\right)$.
	Secondly, singular SVAR models are special in the sense that some
	researcher imposed restrictions are not over-identifying in the sense
	that imposing them does not restrict the feasible covariance structures.
	In Lemma \ref{lem:overidentifying} we provide a condition for checking
	whether the researcher imposed restrictions on $A_{+}$ are over-identifying.
	
	To simplify discussion, we note that vectorizing the (transposed)
	YW equations leads to $\left(I_{n}\otimes\Gamma_{p}\right){\rm vec}\left(A_{+}'\right)={\rm vec}\left(\gamma_{p}'\right)$.
	In \citet{festschriftbdoanderson2010}, the authors choose the first
	linearly independent rows of $\Gamma_{p}$ as a basis of the row space
	(or equivalently column space) of $\Gamma_{p}$ to define a particular
	solution of the YW equations. To fix ideas, consider a $\Gamma_{p}$
	whose first $\left(np-s\right)$ linearly independent rows are selected
	by premultiplying $S_{1}'$ of dimension $\left(\left(np-s\right)\times np\right)$,
	containing only zeros and ones, and denote by $S_{2}'$ the $\left(s\times np\right)$-dimensional
	matrix containing zeros and ones such that $S_{1}'S_{2}=0$. A basis
	of the column space thus consists of the columns of $\Gamma_{p}S_{1}$,
	i.e. the elements $S_{2}'A_{+}'$ are restricted to zero. Restricting
	each column of $A_{+}'$ to be orthogonal to the columns of $S_{2}$
	therefore results in a unique solution of the YW equations, i.e. the
	matrix in brackets in $\left[\begin{smallmatrix}\left(I_{n}\otimes\Gamma_{p}\right)\\
		\left(I_{n}\otimes S_{2}'\right)
	\end{smallmatrix}\right]vec\left(A_{+}'\right)=\left(\begin{smallmatrix}{\rm vec}\left(\gamma_{p}'\right)\\
		0_{n\times1}
	\end{smallmatrix}\right)$ is of full rank. We denote the unique solution of the equation above
	by $\widehat{{\rm vec}\left(A_{+}'\right)}$.
	
	In \citet{chenandersondeistlerfiller2010}, the authors choose the
	minimum norm solution of the YW equations as the particular solution.
	Let $I_{n}\otimes\left[\left(\begin{smallmatrix}V_{1} & V_{2}\end{smallmatrix}\right)\left(\begin{smallmatrix}D_{11} & 0_{(n^{2}p-s)\times s}\\
		0_{s\times(n^{2}p-s)} & 0_{s\times s}
	\end{smallmatrix}\right)\left(\begin{smallmatrix}V_{1}'\\
		V_{2}'
	\end{smallmatrix}\right)\right]$ be the singular value decomposition (SVD)\footnote{$\left(V_{1},V_{2}\right)$ are an orthonormal eigenbasis describing
		the image and the kernel of $\Gamma_{p}$ respectively, and $D_{11}$
		is a diagonal matrix with positive diagonal elements.} of $\left(I_{n}\otimes\Gamma_{p}\right)$ of rank $n^{2}p-ns=n\cdot rk\left(\Gamma_{p}\right)$.
	The particular solution is such that coordinates corresponding to
	the basis vectors $V_{2}$ are set equal to zero. Put differently,
	${\rm vec}\left(A_{+}'\right)$ is required to be orthogonal to the
	columns of $\left(I_{n}\otimes V_{2}\right)$, i.e. $\left[\begin{smallmatrix}\left(I_{n}\otimes\Gamma_{p}\right)\\
		\left(I_{n}\otimes V_{2}'\right)
	\end{smallmatrix}\right]{\rm vec}\left(A_{+}'\right)=\left(\begin{smallmatrix}{\rm vec}\left(\gamma_{p}'\right)\\
		0_{s\times1}
	\end{smallmatrix}\right)$. We denote the unique solution of the equation above by $\widetilde{{\rm vec}\left(A_{+}'\right)}$.
	
	While the coordinate representations $\widehat{{\rm vec}\left(A_{+}'\right)}$
	and $\widetilde{{\rm vec}\left(A_{+}'\right)}$ usually differ, $\left(I_{n}\otimes x_{t-1}'\right)\widehat{{\rm vec}\left(A_{+}'\right)}$
	and $\left(I_{n}\otimes x_{t-1}'\right)\widetilde{{\rm vec}\left(A_{+}'\right)}$
	represent the same projection (component wise on the space spanned
	by the columns of $\Gamma_{p}$ or equivalently on the space spanned
	by the components of $x_{t-1}$). By construction, we have that $span_{C}\left(\Gamma_{p}\right)=span_{C}\left(V_{1}\right)=span_{C}\left(\Gamma_{p}S_{1}\right)$
	and, in particular, that the rank of the projection of $\Gamma_{p}S_{1}$
	on $span_{C}\left(\Gamma_{p}\right)$ is equal to the rank of $\Gamma_{p}$.
	This projection idea can be used to investigate whether researcher
	imposed restrictions on the system parameters are ``true'' restrictions
	(in the sense that they restrict the possible covariance structures
	of the model) and whether the restrictions are sufficient to guarantee
	a unique solution. Let $C_{S}{\rm vec}\left(A_{+}'\right)=0$, where
	$C_{S}\in\mathbb{R}^{r_{S}\times n^{2}p}$ is of full row rank, be
	the researcher imposed restrictions and denote the (right-) kernel
	of $C_{S}$ by $S_{A}\in\mathbb{R}^{n^{2}p\times\left(n^{2}p-r_{S}\right)}$.
	If $span_{C}\left(\left(I_{n}\otimes\Gamma_{p}\right)S_{A}\right)\supseteq span_{C}\left(I_{n}\otimes V_{1}\right)$,
	then the researcher imposed restrictions are not over-identifying
	in the sense that without them the same set of covariance structures
	are feasible. In order to investigate the validity of this inclusion
	of spaces, we define the SVD of 
	\begin{equation}
		\underbrace{\left(I_{n}\otimes\Gamma_{p}\right)S_{A}}_{=n^{2}p\times\left(n^{2}p-r\right)}=\begin{pmatrix}\tilde{U}_{1} & \tilde{U}_{2}\end{pmatrix}\begin{pmatrix}\tilde{D}_{11} & 0\\
			0 & 0_{\tilde{s}\times\tilde{s}}
		\end{pmatrix}\begin{pmatrix}\tilde{V}_{1}'\\
			\tilde{V}_{2}'
		\end{pmatrix}.\label{eq:svd_intersection}
	\end{equation}
	
	If $span_{C}\left(\left(I_{n}\otimes\Gamma_{p}\right)S_{A}\right)\supseteq span_{C}\left(I_{n}\otimes V_{1}\right)$
	holds, then we can express the column space of $\left(I_{n}\otimes V_{1}\right)$
	in terms of the columns of $\left(\left(I_{n}\otimes\Gamma_{p}\right)S_{A}\right)$
	and, in other words, the projection of $\left(I_{n}\otimes V_{1}\right)$
	on the column space of $\left(\left(I_{n}\otimes\Gamma_{p}\right)S_{A}\right)$
	must coincide with $\left(I_{n}\otimes V_{1}\right)$. Expressed in
	terms of SVDs, this leads to
	\begin{lem}
		\label{lem:overidentifying}In the case $A_{0}=I_{n}$, the restrictions
		described by the matrix $C_{S}$ are not over-identifying if and only
		if 
		\begin{equation}
			\left[I_{n^{2}p}-\tilde{U}_{1}\tilde{U}_{1}'\right]\left(I_{n}\otimes V_{1}\right)=0,\label{eq:check_overidentifying}
		\end{equation}
		where $\tilde{U}_{1}$ is obtained from (\ref{eq:svd_intersection}).
		There is a unique solution of the YW equations if and only if the
		right-kernel of $\left(I_{n}\otimes\Gamma_{p}\right)S_{A}$ is trivial.
	\end{lem}
	
	Returning to the general case where $A_{0}$ is not necessarily equal
	to the identity matrix, we will now show that it is in general enough
	to impose as many restrictions as there are basis vectors in the kernel
	of $\left(I_{n}\otimes\Gamma_{p}\right)$. Notice that $C_{S}vec\left(A'_{+}\right)=\left[C_{S}\left(A_{0}\otimes I_{np}\right)\right]vec\left(\bar{A}'_{+}\right)$
	such that for given $A_{0}$, the restrictions on the parameters $\bar{A}_{+}$
	can be obtained straight-forwardly from the ones on $A_{+}$.
	
	In order to provide some intuition for the following result, we consider
	a quite special example where it is not sufficient to count the number
	of restrictions for deducing identifiability of the system parameters.
	Consider $\Gamma_{p}=\left(\begin{smallmatrix}1 & 0 & 0\\
		0 & 1 & 0\\
		0 & 0 & 0
	\end{smallmatrix}\right)$ and $C_{S}=I_{3}\otimes\left(0,1,0\right)$, such that $S_{A}=\left(I_{3}\otimes\left(\begin{smallmatrix}1 & 0\\
		0 & 0\\
		0 & 1
	\end{smallmatrix}\right)\right)$ and $\left(I_{n}\otimes\Gamma_{p}\right)S_{A}=\left(I_{3}\otimes\left(\begin{smallmatrix}1 & 0\\
		0 & 0\\
		0 & 0
	\end{smallmatrix}\right)\right)$. Even though the order condition (that the rank deficiency of $\left(I_{n}\otimes\Gamma_{p}\right)$
	is equal to the number of restrictions) is satisfied, they are not
	sufficient for obtaining a unique solution of the YW equations. Indeed,
	$\left[I_{n^{2}p}-\tilde{U}_{1}\tilde{U}_{1}'\right]\left(I_{n}\otimes V_{1}\right)=\left(I_{3}\otimes\left(\begin{smallmatrix}0 & 0 & 0\\
		0 & 1 & 0\\
		0 & 0 & 0
	\end{smallmatrix}\right)\right)\left(I_{3}\otimes\left(\begin{smallmatrix}1 & 0\\
		0 & 1\\
		0 & 0
	\end{smallmatrix}\right)\right)\neq0$ and the right-kernel of $\left(I_{n}\otimes\Gamma_{p}\right)S_{A}=\left(I_{3}\otimes\left(\begin{smallmatrix}1 & 0\\
		0 & 0\\
		0 & 0
	\end{smallmatrix}\right)\right)$ is non-empty. The non-generic nature of this example is summarized
	in
	\begin{prop}
		\label{prop:Aplus}Let $C_{S}\in\mathbb{R}^{ns\times n^{2}p}$ be
		of full row rank and let $\Gamma_{p}$ be singular with rank deficiency
		equal to $s$. The set of restrictions $\left\{ C_{S}\in\mathbb{R}^{ns\times n^{2}p}\,|\,\eqref{eq:check_overidentifying}\text{ does not hold}\right\} $
		is of Lebesgue measure zero in $\mathbb{R}^{ns\times n^{2}p}$. A
		generic, randomly chosen restriction $C_{S}$ can thus be used to
		obtain a unique solution of the system of equations $\left[\begin{smallmatrix}\left(I_{n}\otimes\Gamma_{p}\right)\\
			C_{S}
		\end{smallmatrix}\right]{\rm vec}\left(A_{+}'\right)=\left(\begin{smallmatrix}{\rm vec}\left(\gamma_{p}'\right)\\
			0_{s\times1}
		\end{smallmatrix}\right)$ and the system parameters are globally identified.
	\end{prop}
	
	\textcolor{red}{}Notice, however, that it is not possible to solve
	the identifiability problem for $A_{+}$ by restricting the transfer
	function $a(z)^{-1}b$ (e.g., by restricting the long-run coefficients
	in $k(1)=a(1)^{-1}b$). Since two observationally equivalent pairs
	$\left(a^{(1)}(z),b^{(1)}\right)$ and $\left(a^{(2)}(z),b^{(2)}\right)$
	have by definition the same transfer function, restricting $a(z)^{-1}b$
	directly has the effect of either excluding the whole equivalence
	class or not providing additional information for distinguishing different
	pairs $\left(a(z),b\right)$ with the same transfer function.

	\section{Conclusion}
	
	In this article, we generalize the well-known identifiability results
	for SVAR models to the case of a singular innovation covariance matrix.
	The first main difference to the regular case is that the restrictions
	on the noise parameters $(A_{0},B)$ might contradict the singularity
	of the innovation covariance matrix. Moreover, the researcher imposed
	restrictions might already be contained in the restrictions implied
	by the singularity of the innovation covariance matrix and therefore
	do not have any further ``identifying effect''. The second main
	difference pertains mainly to restrictions on the structural system
	parameters $A_{+}$. We provide conditions under which the researcher
	imposed restrictions on these parameters are over-identifying and
	show that underidentification can be considered an unusual case when
	the rank deficiency coincides with the number of restrictions. 
	
	\section{Acknowledgements}
	
	Financial support by the Research Funds of the University of Helsinki
	as well as by funds of the Oesterreichische Nationalbank (Austrian
	Central Bank, Anniversary Fund, project number: 17646) is gratefully
	acknowledged.
	
	\section{Data Availability Statement}
	
	There is no data involved in this study.
	
	\pagebreak{}

	\newpage{}
	
	\selectlanguage{british}%
	
	\appendix
	\setcounter{page}{1}
	\renewcommand{\thepage}{A-\arabic{page}} 
	\newgeometry{left=1cm,right=1cm,top=2cm,bottom=2cm}
	
	\selectlanguage{english}%

	\section{Proof of Lemma \ref{lem:rest_affine_compatible}}
	
	We write $C_{N}=\left(\begin{smallmatrix}C_{A_{0}} & 0_{r_{A_{0}}\times nq}\\
		0_{r_{B}\times n^{2}} & C_{B}
	\end{smallmatrix}\right)$ as orthogonal sum, i.e. 
	\[
	C_{N}=Proj_{R}(C_{N}|\mathcal{N})+\underbrace{\left(C_{N}-Proj_{R}(C_{N}|\mathcal{N})\right)}_{=M},
	\]
	and substitute it into equation (\ref{eq:restr_B_affine}) such that
	$Proj_{R}(C_{N}|\mathcal{N}){\rm vec}(A_{0},B)+M{\rm vec}(A_{0},B)=c_{N}.$
	In order to fulfill the singularity restrictions of $\Sigma_{u}$,
	equation (\ref{eq:restr_B_singul}) needs to hold. Elementary calculations
	show that $\mathcal{N}{\rm vec}\left(A_{0},B\right)=0$ under (\ref{eq:restr_B_singul})
	which implies $Proj_{R}(C_{N}|\mathcal{N}){\rm vec}(A_{0},B)=0$ because
	$Proj_{R}(C_{N}|\mathcal{N})$ projects $C_{N}$ onto $span_{R}\left(\mathcal{N}\right)$.
	The system of equations $M{\rm vec}(A_{0},B)=c_{N}$ has a solution
	if and only if ${\rm rk}(M)={\rm rk}\begin{pmatrix}M & c_{N}\end{pmatrix}$.
	
	\section{Proof of Proposition \ref{prop:BmodelLocalIdent}}
	
	Consider the following system of equations:
	\begin{align*}
		\varphi_{1}({\rm vec}(A_{0},B)):={\rm vech}\left(A_{0}^{-1}BB'A_{0}'^{-1}\right)-{\rm vech}\left(\Sigma_{u}\right) & =0,\\
		\varphi_{2}({\rm vec}(A_{0},B)):=C_{N}{\rm vec}(A_{0},B)-c_{N}=\left(\begin{smallmatrix}C_{A_{0}} & 0_{r_{A_{0}}\times nq}\\
			0_{r_{B}\times n^{2}} & C_{B}
		\end{smallmatrix}\right) & {\rm vec}(A_{0},B)-c_{N}=0.
	\end{align*}
	Following \citet[Theorem 6]{Rothenberg71}, the equations $\varphi({\rm vec}(A_{0},B))=\left(\begin{smallmatrix}\varphi_{1}({\rm vec}(A_{0},B))\\
		\varphi_{2}({\rm vec}(A_{0},B))
	\end{smallmatrix}\right)=0\in\mathbb{R}^{\left(\frac{n(n+1)}{2}+r_{N}\right)\times1}$ have a unique solution in an open set around ${\rm vec}(A_{0},B)\in\mathbb{R}^{\left(n^{2}+nq\right)\times1}$
	if the $\left(\frac{n(n+1)}{2}+r_{N}\right)\times\left(n^{2}+nq\right)$
	dimensional matrix $\frac{\partial\varphi}{\partial{\rm vec}(A_{0},B)'}$
	has full column rank $n^{2}+nq$. Note that $\varphi({\rm vec}(A_{0},B))=0$
	holds if and only if $c_{N}$ is in the image of $M=C_{N}-Proj_{R}(C_{N}|\mathcal{N})$
	according to Lemma \ref{lem:rest_affine_compatible}. The matrix $\frac{\partial\varphi}{\partial{\rm vec}(A_{0},B)'}$
	can be calculated using standard rules for matrix differentiation
	\citep{luet_mat96} as
	\begin{align*}
		\frac{\partial{\rm \varphi_{1}}}{\partial{\rm vec}(A_{0})'}({\rm vec}(A_{0},B)) & =D_{n}^{+}\frac{\partial{\rm vec}\left(A_{0}^{-1}BB'A_{0}'^{-1}\right)}{\partial{\rm vec}(A_{0}^{-1}B)'}\frac{\partial{\rm vec}\left(A_{0}^{-1}B\right)}{\partial{\rm vec}(A_{0})'}\\
		& =D_{n}^{+}\left((I\otimes A_{0}^{-1}B)K_{nq}\frac{\partial{\rm vec}\left(A_{0}^{-1}B\right)}{\partial{\rm vec}(A_{0}^{-1}B)'}+(A_{0}^{-1}B\otimes I)\frac{\partial{\rm vec}(A_{0}^{-1}B)}{\partial{\rm vec}(A_{0}^{-1}B)'}\right)\frac{\partial{\rm vec}\left(A_{0}^{-1}B\right)}{\partial{\rm vec}(A_{0})'}\\
		& =-2D_{n}^{+}(\Sigma_{u}\otimes A_{0}^{-1}),
	\end{align*}
	
	\begin{align*}
		\frac{\partial{\rm \varphi_{1}}}{\partial{\rm vec}(B)'}({\rm vec}(A_{0},B)) & =\frac{\partial{\rm vech}(A_{0}^{-1}BB'A_{0}'^{-1})}{\partial{\rm vec}(B)'}=D_{n}^{+}\frac{\partial{\rm vec}\left(A_{0}^{-1}BB'A_{0}'^{-1}\right)}{\partial{\rm vec}(A_{0}^{-1}B)'}\frac{\partial{\rm vec}\left(A_{0}^{-1}B\right)}{\partial{\rm vec}(B)'}\\
		& =D_{n}^{+}\left((I\otimes A_{0}^{-1}B)K_{nq}\frac{\partial{\rm vec}\left(A_{0}^{-1}B\right)}{\partial{\rm vec}(A_{0}^{-1}B)'}+(A_{0}^{-1}B\otimes I)\frac{\partial{\rm vec}(A_{0}^{-1}B)}{\partial{\rm vec}(A_{0}^{-1}B)'}\right)A_{0}^{-1}\\
		& =2D_{n}^{+}(A_{0}^{-1}B\otimes A_{0}^{-1}),
	\end{align*}
	and $\frac{\partial\varphi_{2}}{\partial{\rm vec}(A_{0},B)'}({\rm vec}(A_{0},B))=C_{N}.$
	Here, $D_{n}^{+}$ is the pseudo-inverse of the duplication matrix
	$D_{n}$ which fulfills $D_{n}{\rm vech}(A)={\rm vec}(A)$ for a matrix
	$A\in\mathbb{R}^{n\times n}$, and $K_{nm}\in\mathbb{R}^{nm\times nm}$
	is a commutation matrix such that ${\rm vec}(B')=K_{nm}{\rm vec}(B)$
	for $B\in\mathbb{R}^{n\times m}$, see e.g. \citet[page 662 and 663]{luet05}.

	\section{Proof of Proposition \ref{prop:Aplus}}
	
	Let $S_{A}$ of dimension $\left(n^{2}p\times n(np-s)\right)$ denote
	the matrix obtained as the orthogonal complement of $C_{A}$. Since
	$\left(I_{n}\otimes\Gamma_{p}\right)=\left(I_{n}\otimes\left(\begin{smallmatrix}V_{1} & V_{2}\end{smallmatrix}\right)\left(\begin{smallmatrix}D_{11} & 0_{(n^{2}p-s)\times s}\\
		0_{s\times(n^{2}p-s)} & 0_{s\times s}
	\end{smallmatrix}\right)\left(\begin{smallmatrix}V_{1}'\\
		V_{2}'
	\end{smallmatrix}\right)\right)$, it is obvious that $\left(I_{n}\otimes\Gamma_{p}\right)S_{A}$ does
	not have full rank if and only if $\left(I_{n}\otimes V_{1}'\right)S_{A}$
	is of reduced rank (smaller than $n^{2}p-ns$). For given $\Gamma_{p}$,
	the elements in the matrix of restrictions $C_{A}$ (and therefore
	also the ones in $S_{A}$) are free (up to the requirement that the
	rows of $C_{A}$ be linearly independent). The determinant $\det\left(\left(I_{n}\otimes V_{1}'\right)S_{A}\right)$
	is thus a multivariate polynomial in the elements of $S_{A}$. This
	determinant is either identically zero or zero only on a set of Lebesgue
	measure zero. Since for $S_{A}=\left(I_{n}\otimes V_{1}\right)$ the
	determinant is equal to one, the determinant is not identically zero.
	
\newpage

\section{Supplementary Appendix: Details for Rational Expectations Model}

This supplementary appendix analyses a version of the New Keynesian monetary model
with a ``supply-shifting'' and a monetary policy shock. The first
section introduces the economic model and states the structural
difference equation involving conditional expectations of future
endogenous variables. The second section verifies that for the parameter
values used in the main article, there are two unstable eigenvalues and
that the existence and uniqueness condition in
\href{https://doi.org/10.1023/A:1020517101123}{Sims (2001)} is
satisfied.

\hypertarget{description-of-rational-expectations-model}{%
\subsection{Description of Rational Expectations
Model}\label{description-of-rational-expectations-model}}

We consider the model

\[
\begin{array}{rl}
    \pi_{t} & =\beta\mathbb{E}_{t}\left(\pi_{t+1}\right)+\kappa x_{t}+\varepsilon_{t}^{\pi}\\
x_{t} & =\mathbb{E}_{t}\left(x_{t+1}\right)-\tau\left(R_{t}-\mathbb{E}_{t}\left(\pi_{t+1}\right)\right)\\
R_{t} & =\phi\mathbb{E}_{t}\left(\pi_{t+1}\right)+\varepsilon_{t}^{R}.
\end{array}
\]

where \(\pi_t\), \(x_t\) and \(R_t\) are log-deviations from the steady
state of inflation, output gap and the interest rate. The conditional
expectations are to be understood as linear projections on the space
spanned by present and past components of the uncorrelated shocks
\(\varepsilon_{t}^{\pi}\) and \(\varepsilon_{t}^{R}\) which are white
noise processes (whose variance is normalised to one for the sake of
simplicity). The parameters of the model are the subjective time
preference factor \(\beta\in\left(0,1\right)\), \(\phi\geq0\) the
elasticity of the interest response of the central bank, and the slope
parameters \(\kappa\) and \(\tau\).

\hypertarget{rewriting-the-structural-differnce-equation-by-introducting-new-variables}{%
\subsubsection{Rewriting the Structural Differnce Equation by
Introducting New
Variables}\label{rewriting-the-structural-differnce-equation-by-introducting-new-variables}}

Replacing the conditional expectations with new variables
\(\mathbb{E}_{t} \left( x_{t+1} \right) = \xi^{x}_t\) and
\(\mathbb{E}_{t} \left( \pi_{t+1} \right) = \xi^{\pi}_t\), adding
equations

\[
\begin{array}{rl}
  x_{t} & = \xi^{x}_{t-1} + \eta^{x}_t\\
  \pi_{t} & = \xi^{\pi}_{t-1} + \eta^{\pi}_t
\end{array}
\]

and replacing \(x_t\) and \(\pi_t\) in the three equations above, leads
to the following system in the canoncial form of
\href{https://doi.org/10.1023/A:1020517101123}{Sims (2001)}

\[
\begin{pmatrix}
    \beta & 0 & 0 & 0 & 0\\
    \tau & 1 & -\tau & 0 &  0\\
    -\phi & 0 & 1 & 0 & 0 \\
    0 & 0 & 0 & 1 & 0 \\
    0 & 0 & 0 & 0 & 1
\end{pmatrix}
\begin{pmatrix}
    \xi_{t}^{\pi}\\
    \xi_{t}^{x}\\
    R_{t} \\
    \pi_t \\
    x_t
\end{pmatrix} =
\begin{pmatrix}
    1 & -\kappa & 0 & 0 & 0\\
    0 & 1 & 0 & 0 & 0 \\
    0 & 0 & 0 & 0 & 0 \\
    1 & 0 & 0 & 0 & 0 \\
    0 & 1 & 0 & 0 & 0 
\end{pmatrix}
\begin{pmatrix}
    \xi_{t-1}^{\pi}\\
    \xi_{t-1}^{x}\\
    R_{t-1} \\
    \pi_{t-1} \\
    x_{t-1}
\end{pmatrix} + 
\begin{pmatrix}
    1 & -\kappa\\
    0 & 1 \\
    0 & 0 \\
    1 & 0 \\
    0 & 1
\end{pmatrix}
\begin{pmatrix}
    \eta_{t}^{\pi}\\
    \eta_{t}^{x}
\end{pmatrix} +
\begin{pmatrix}
    1 & 0\\
    0 & 0\\
    0 & 1\\
    0 & 0 \\
    0 & 0 
\end{pmatrix}
\begin{pmatrix}
\varepsilon_{t}^{\pi} \\
\varepsilon_{t}^{R} 
\end{pmatrix}
\] where the matrices correspond to
\(\left( \Gamma_0, \Gamma_1, \Pi, \Psi \right)\) in Sims' notation.

\hypertarget{obtaining-the-causal-stationary-solution}{%
\subsection{Obtaining the Causal Stationary
Solution}\label{obtaining-the-causal-stationary-solution}}

Here, we show how to obtain the unique causal stationary solution

\[
\left(\begin{smallmatrix}R_{t}\\
\pi_{t}\\
x_{t}
\end{smallmatrix}\right)=\left(\begin{smallmatrix}1 & 0\\
-\kappa\tau & 1\\
-\tau & 0
\end{smallmatrix}\right)\left(\begin{smallmatrix}\varepsilon_{t}^{R}\\
\varepsilon_{t}^{\pi}
\end{smallmatrix}\right)
\]

of the DSGE model described above. Since the last two variables
\(\pi_t\) and \(x_t\) do not enter the first three equations, we first
derive a causal stationary solution of the the sub-system

\[
\begin{pmatrix}
    \beta & 0 & 0 \\
    \tau & 1 & -\tau \\
    -\phi & 0 & 1  
\end{pmatrix}
\begin{pmatrix}
    \xi_{t}^{\pi}\\
    \xi_{t}^{x}\\
    R_{t} \\
\end{pmatrix} =
\begin{pmatrix}
    1 & -\kappa & 0 \\
    0 & 1 & 0  \\
    0 & 0 & 0  \\
\end{pmatrix}
\begin{pmatrix}
    \xi_{t-1}^{\pi}\\
    \xi_{t-1}^{x}\\
    R_{t-1} 
\end{pmatrix} + 
\begin{pmatrix}
    1 & -\kappa\\
    0 & 1 \\
    0 & 0 \\
\end{pmatrix}
\begin{pmatrix}
    \eta_{t}^{\pi}\\
    \eta_{t}^{x}
\end{pmatrix} +
\begin{pmatrix}
    1 & 0\\
    0 & 0\\
    0 & 1\\
\end{pmatrix}
\begin{pmatrix}
\varepsilon_{t}^{\pi} \\
\varepsilon_{t}^{R} 
\end{pmatrix}
\]

of the system in canonical form above, where the matrices are now
denoted by \(\left( \Gamma_0^s, \Gamma_1^s, \Pi^s, \Psi^s \right)\). To
this end, we consider the eigenvalue decomposition
\((\Gamma_0^s)^{-1} \Gamma_1^s = V \Lambda V'\) and pre-multiply the
system above by \(V^{-1} (\Gamma_0^s)^{-1}\) to obtain

\[
V^{-1}
\begin{pmatrix}
    \xi_{t}^{\pi}\\
    \xi_{t}^{x}\\
    R_{t} \\
\end{pmatrix} =
\begin{pmatrix}
    \lambda_1 & 0 & 0 \\
    0 & \lambda_2 & 0  \\
    0 & 0 & 0  \\
\end{pmatrix}
V^{-1}
\begin{pmatrix}
    \xi_{t-1}^{\pi}\\
    \xi_{t-1}^{x}\\
    R_{t-1} 
\end{pmatrix} + 
V^{-1} (\Gamma_0^s)^{-1}
\begin{pmatrix}
    1 & -\kappa\\
    0 & 1 \\
    0 & 0 \\
\end{pmatrix}
\begin{pmatrix}
    \eta_{t}^{\pi}\\
    \eta_{t}^{x}
\end{pmatrix} +
V^{-1} (\Gamma_0^s)^{-1}
\begin{pmatrix}
    1 & 0\\
    0 & 0\\
    0 & 1\\
\end{pmatrix}
\begin{pmatrix}
\varepsilon_{t}^{\pi} \\
\varepsilon_{t}^{R} 
\end{pmatrix}
\] Note that the parameter values \((\beta,\phi,\tau,\kappa)\) are
chosen such that there are two unstable eigenvalues,
i.e.~\(|\lambda_1| > |\lambda_2| > 1\). When the ``endogenous forecast
errors'' (which are under the control of the economic agents) are such
that they offset the exogenous driving forces in the ``unstable
equations'', i.e.~when

\[
\begin{pmatrix}
    \eta_{t}^{\pi}\\
    \eta_{t}^{x}
\end{pmatrix} = 
\Pi_u^{-1} \Psi_u \begin{pmatrix}
\varepsilon_{t}^{\pi} \\
\varepsilon_{t}^{R} 
\end{pmatrix},
\] where \(\Pi_u\) contains the first two rows of
\(V^{-1} (\Gamma_0^s)^{-1} \begin{pmatrix} 1 & -\kappa\\ 0 & 1 \\ 0 & 0 \\ \end{pmatrix}\)
and \(\Psi_u\) contains the first two rows of
\(V^{-1} (\Gamma_0^s)^{-1} \begin{pmatrix} 1 & 0\\ 0 & 0\\ 0 & 1\\ \end{pmatrix}\),
we obtain that \(\xi_t^{\pi} = \xi_t^{x} = 0\) for all \(t\).
Transforming back to the original variables results in a causal
stationary solution of the structural difference equation involving
conditional expectations.

With the following code, we verify that is indeed
\(\Pi_u^{-1} \Psi_u = \begin{pmatrix} 1 & - \kappa \tau \\ 0 & - \tau \end{pmatrix}\).

\begin{Shaded}
\begin{Highlighting}[]
\ImportTok{from}\NormalTok{ sympy }\ImportTok{import} \OperatorTok{*}

\NormalTok{init\_printing(use\_unicode}\OperatorTok{=}\VariableTok{False}\NormalTok{)}

\CommentTok{\# Define deep parameters}
\NormalTok{b, t, f, k }\OperatorTok{=}\NormalTok{ symbols(}\StringTok{"b t f k"}\NormalTok{)}
\NormalTok{l }\OperatorTok{=}\NormalTok{ symbols(}\StringTok{"l"}\NormalTok{)  }\CommentTok{\# eigenvalue variable}

\CommentTok{\# Corresponds to \textbackslash{}Gamma\_\{0\} in Sims\textquotesingle{} notation}
\NormalTok{g0 }\OperatorTok{=}\NormalTok{ Matrix([[b, }\DecValTok{0}\NormalTok{, }\DecValTok{0}\NormalTok{], }
\NormalTok{             [t, }\DecValTok{1}\NormalTok{, }\OperatorTok{{-}}\NormalTok{t], }
\NormalTok{             [}\OperatorTok{{-}}\NormalTok{f, }\DecValTok{0}\NormalTok{, }\DecValTok{1}\NormalTok{]])}

\CommentTok{\# Corresponds to \textbackslash{}Gamma\_\{1\} in Sims\textquotesingle{} notation}
\NormalTok{g1 }\OperatorTok{=}\NormalTok{ Matrix([[}\DecValTok{1}\NormalTok{, }\OperatorTok{{-}}\NormalTok{k, }\DecValTok{0}\NormalTok{], }
\NormalTok{             [}\DecValTok{0}\NormalTok{, }\DecValTok{1}\NormalTok{, }\DecValTok{0}\NormalTok{], }
\NormalTok{             [}\DecValTok{0}\NormalTok{, }\DecValTok{0}\NormalTok{, }\DecValTok{0}\NormalTok{]])}

\CommentTok{\# Corresponds to \textbackslash{}Pi in Sims\textquotesingle{} notation}
\NormalTok{pp }\OperatorTok{=}\NormalTok{ Matrix([[}\DecValTok{1}\NormalTok{, }\OperatorTok{{-}}\NormalTok{k], }
\NormalTok{             [}\DecValTok{0}\NormalTok{, }\DecValTok{1}\NormalTok{], }
\NormalTok{             [}\DecValTok{0}\NormalTok{, }\DecValTok{0}\NormalTok{]])}
             
\CommentTok{\# Corresponds to \textbackslash{}Psi in Sims\textquotesingle{} notation}
\NormalTok{ff }\OperatorTok{=}\NormalTok{ Matrix([[}\OperatorTok{{-}}\DecValTok{1}\NormalTok{, }\DecValTok{0}\NormalTok{], }
\NormalTok{             [}\DecValTok{0}\NormalTok{, }\DecValTok{0}\NormalTok{], }
\NormalTok{             [}\DecValTok{0}\NormalTok{, }\DecValTok{1}\NormalTok{]])}

\CommentTok{\# Premultiply the inverse of \textbackslash{}Gamma\_0}
\NormalTok{gg1 }\OperatorTok{=}\NormalTok{ g0.inv() }\OperatorTok{*}\NormalTok{ g1}
\NormalTok{gpp }\OperatorTok{=}\NormalTok{ g0.inv() }\OperatorTok{*}\NormalTok{ pp}
\NormalTok{gff }\OperatorTok{=}\NormalTok{ g0.inv() }\OperatorTok{*}\NormalTok{ ff}

\CommentTok{\# Replacement of symbols with specifice parameter values}
\NormalTok{par }\OperatorTok{=}\NormalTok{ [(b, }\DecValTok{4} \OperatorTok{/} \DecValTok{5}\NormalTok{), (f, }\DecValTok{39} \OperatorTok{/} \DecValTok{38}\NormalTok{), (t, }\DecValTok{3} \OperatorTok{/} \DecValTok{4}\NormalTok{), (k, }\DecValTok{1} \OperatorTok{/} \DecValTok{2}\NormalTok{)]}

\CommentTok{\# Calculate the eigenvectors}
\NormalTok{evec }\OperatorTok{=}\NormalTok{ gg1.eigenvects(simplify}\OperatorTok{=}\StringTok{"True"}\NormalTok{)}

\CommentTok{\# Concatenating the eigenvectors into a matrix}
\NormalTok{M }\OperatorTok{=}\NormalTok{ evec[}\DecValTok{1}\NormalTok{][}\DecValTok{2}\NormalTok{][}\DecValTok{0}\NormalTok{].col\_insert(}\DecValTok{1}\NormalTok{, evec[}\DecValTok{2}\NormalTok{][}\DecValTok{2}\NormalTok{][}\DecValTok{0}\NormalTok{]).col\_insert(}\DecValTok{2}\NormalTok{, evec[}\DecValTok{0}\NormalTok{][}\DecValTok{2}\NormalTok{][}\DecValTok{0}\NormalTok{])}

\CommentTok{\# Inverse of eigenvector matrix}
\NormalTok{Mi }\OperatorTok{=}\NormalTok{ M.inv()}

\CommentTok{\# Calculate the transformed \textbackslash{}Pi and \textbackslash{}Psi matrix pertaining to \textquotesingle{}unstable coordinates\textquotesingle{}}
\NormalTok{gppu }\OperatorTok{=}\NormalTok{ simplify(Mi.row([}\DecValTok{0}\NormalTok{, }\DecValTok{1}\NormalTok{]) }\OperatorTok{*}\NormalTok{ gpp)}
\NormalTok{gffu }\OperatorTok{=}\NormalTok{ simplify(Mi.row([}\DecValTok{0}\NormalTok{, }\DecValTok{1}\NormalTok{]) }\OperatorTok{*}\NormalTok{ gff)}
\end{Highlighting}
\end{Shaded}

\hypertarget{evaluating-identifiability-of-b}{%
\subsection{Evaluating Identifiability of
B}\label{evaluating-identifiability-of-b}}

In order to apply Proposition 1, we need to calculate

\begin{itemize}
\tightlist
\item
  the perpendicular of the projection of \(C_{B}\) on the row-span of
  \(\left(I_{2}\otimes L\right)\) for
  \(L=\begin{pmatrix}\tau & 0 & 1\end{pmatrix}\), i.e.~calculate the
  Matrix \(M\) in Lemma 1 and Proposition 1, and
\item
  evaluate the rank of the matrix involving the pseudo-inverse of the
  duplication matrix \(D_3\) in Proposition 1.
\end{itemize}

First, we calculate \(M\). Subsequently, we verify that
\(rank( M ) = rank( M \, | \, c_B)\). Last, we check the rank of the
matrix
\(\left(\begin{smallmatrix}2D_{n}^{+}(B\otimes I_{n})\\ C_{B} \end{smallmatrix}\right)\)
in Proposition 1.

\begin{Shaded}
\begin{Highlighting}[]
\CommentTok{\# Matrix of restrictions on noise parameters in vec(B)}
\NormalTok{Cb }\OperatorTok{=}\NormalTok{  Matrix([[}\DecValTok{1}\NormalTok{, }\DecValTok{0}\NormalTok{, }\DecValTok{0}\NormalTok{, }\DecValTok{0}\NormalTok{, }\DecValTok{0}\NormalTok{, }\DecValTok{0}\NormalTok{], }
\NormalTok{              [}\DecValTok{0}\NormalTok{, }\DecValTok{0}\NormalTok{, }\DecValTok{0}\NormalTok{, }\DecValTok{1}\NormalTok{, }\DecValTok{0}\NormalTok{ ,}\DecValTok{0}\NormalTok{], }
\NormalTok{              [}\DecValTok{0}\NormalTok{, }\DecValTok{0}\NormalTok{, }\DecValTok{0}\NormalTok{, }\DecValTok{0}\NormalTok{, }\DecValTok{1}\NormalTok{, }\DecValTok{0}\NormalTok{], }
\NormalTok{              [}\DecValTok{0}\NormalTok{, }\DecValTok{0}\NormalTok{, }\DecValTok{0}\NormalTok{, }\DecValTok{0}\NormalTok{, }\DecValTok{0}\NormalTok{, }\DecValTok{1}\NormalTok{]])}

\ImportTok{import}\NormalTok{ numpy }\ImportTok{as}\NormalTok{ np}

\CommentTok{\# 6{-}dimensional identity matrix}
\NormalTok{I6 }\OperatorTok{=}\NormalTok{ Matrix(np.eye(}\DecValTok{6}\NormalTok{))}

\CommentTok{\# Kronecker product of I\_2 and (the transpose of) L, and their crossproduct}
\NormalTok{I2\_otimes\_L\_transpose }\OperatorTok{=}\NormalTok{ Matrix([[t, }\DecValTok{0}\NormalTok{], }
\NormalTok{                                [}\DecValTok{0}\NormalTok{, }\DecValTok{0}\NormalTok{], }
\NormalTok{                                [}\DecValTok{1}\NormalTok{, }\DecValTok{0}\NormalTok{], }
\NormalTok{                                [}\DecValTok{0}\NormalTok{, t], }
\NormalTok{                                [}\DecValTok{0}\NormalTok{, }\DecValTok{0}\NormalTok{], }
\NormalTok{                                [}\DecValTok{0}\NormalTok{, }\DecValTok{1}\NormalTok{]])}
\NormalTok{I2\_otimes\_L }\OperatorTok{=}\NormalTok{ Matrix(np.transpose(I2\_otimes\_L\_transpose))}
\NormalTok{cp }\OperatorTok{=}\NormalTok{ I2\_otimes\_L}\OperatorTok{*}\NormalTok{I2\_otimes\_L\_transpose}

\CommentTok{\# Calculating and simplifying M}
\NormalTok{M }\OperatorTok{=}\NormalTok{ Cb }\OperatorTok{*}\NormalTok{ (I6 }\OperatorTok{{-}}\NormalTok{ I2\_otimes\_L\_transpose }\OperatorTok{*}\NormalTok{ cp.inv() }\OperatorTok{*}\NormalTok{ I2\_otimes\_L)}
\NormalTok{simplify(M)}
\end{Highlighting}
\end{Shaded}

\begin{verbatim}
## [ 1.0        -t                        ]
## [------  0  ------    0      0     0   ]
## [ 2          2                         ]
## [t  + 1     t  + 1                     ]
## [                                      ]
## [                    1.0          -t   ]
## [  0     0    0     ------   0   ------]
## [                    2            2    ]
## [                   t  + 1       t  + 1]
## [                                      ]
## [  0     0    0       0     1.0    0   ]
## [                                      ]
## [                                     2]
## [                    -t          1.0*t ]
## [  0     0    0     ------   0   ------]
## [                    2            2    ]
## [                   t  + 1       t  + 1]
\end{verbatim}

For specific parameter values
\((\beta,\phi,\tau,\kappa)=\left(\frac{4}{5},\frac{39}{38},\frac{3}{4},\frac{1}{2}\right)\),
we check whether the rank condition of Lemma 1 is satisfied. Below, we
verify that \(rank( M ) = 3\) and that \(rank( M \, | \, c_B) = 3\) such
that we may apply Proposition 1.

\begin{Shaded}
\begin{Highlighting}[]
\DocumentationTok{\#\# right hand side c\_b:}
\NormalTok{cb }\OtherTok{=} \FunctionTok{matrix}\NormalTok{(}\FunctionTok{c}\NormalTok{(}\DecValTok{1}\NormalTok{, }\DecValTok{0}\NormalTok{, }\DecValTok{1}\NormalTok{, }\DecValTok{0}\NormalTok{))}
\NormalTok{CB }\OtherTok{=} \FunctionTok{matrix}\NormalTok{(}\FunctionTok{c}\NormalTok{(}\DecValTok{1}\NormalTok{,}\FunctionTok{rep}\NormalTok{(}\DecValTok{0}\NormalTok{, }\DecValTok{5}\NormalTok{), }
              \FunctionTok{rep}\NormalTok{(}\DecValTok{0}\NormalTok{, }\DecValTok{3}\NormalTok{), }\DecValTok{1}\NormalTok{, }\DecValTok{0}\NormalTok{, }\DecValTok{0}\NormalTok{, }
              \FunctionTok{rep}\NormalTok{(}\DecValTok{0}\NormalTok{, }\DecValTok{4}\NormalTok{), }\DecValTok{1}\NormalTok{, }\DecValTok{0}\NormalTok{, }
              \FunctionTok{rep}\NormalTok{(}\DecValTok{0}\NormalTok{, }\DecValTok{5}\NormalTok{), }\DecValTok{1}\NormalTok{), }
            \AttributeTok{byrow =} \ConstantTok{TRUE}\NormalTok{, }\AttributeTok{ncol =} \DecValTok{6}\NormalTok{)}

\DocumentationTok{\#\# M matrices}

\NormalTok{m }\OtherTok{\textless{}{-}} \ControlFlowTok{function}\NormalTok{(t) \{}
\NormalTok{  t2p }\OtherTok{\textless{}{-}}\NormalTok{ t}\SpecialCharTok{\^{}}\DecValTok{2}\SpecialCharTok{+}\DecValTok{1} 
  \FunctionTok{matrix}\NormalTok{(}\FunctionTok{c}\NormalTok{(}\DecValTok{1}\SpecialCharTok{{-}}\NormalTok{(t}\SpecialCharTok{\^{}}\DecValTok{2}\SpecialCharTok{/}\NormalTok{t2p), }\DecValTok{0}\NormalTok{, }\SpecialCharTok{{-}}\NormalTok{t}\SpecialCharTok{/}\NormalTok{(t2p), }\DecValTok{0}\NormalTok{, }\DecValTok{0}\NormalTok{, }\DecValTok{0}\NormalTok{,}
           \DecValTok{0}\NormalTok{, }\DecValTok{0}\NormalTok{, }\DecValTok{0}\NormalTok{, }\DecValTok{1}\SpecialCharTok{{-}}\NormalTok{(t}\SpecialCharTok{\^{}}\DecValTok{2}\SpecialCharTok{/}\NormalTok{t2p), }\DecValTok{0}\NormalTok{, }\SpecialCharTok{{-}}\NormalTok{t}\SpecialCharTok{/}\NormalTok{(t2p), }
           \DecValTok{0}\NormalTok{, }\DecValTok{0}\NormalTok{, }\DecValTok{0}\NormalTok{, }\DecValTok{0}\NormalTok{, }\DecValTok{1}\NormalTok{, }\DecValTok{0}\NormalTok{, }
           \DecValTok{0}\NormalTok{, }\DecValTok{0}\NormalTok{, }\DecValTok{0}\NormalTok{, }\SpecialCharTok{{-}}\NormalTok{t}\SpecialCharTok{/}\NormalTok{(t2p), }\DecValTok{0}\NormalTok{, }\DecValTok{1}\SpecialCharTok{{-}}\NormalTok{(}\DecValTok{1}\SpecialCharTok{/}\NormalTok{t2p)), }
         \AttributeTok{byrow=}\ConstantTok{TRUE}\NormalTok{, }\AttributeTok{ncol=}\DecValTok{6}\NormalTok{)}
\NormalTok{\} }


\FunctionTok{qr}\NormalTok{(}\FunctionTok{m}\NormalTok{(}\FloatTok{0.75}\NormalTok{))}\SpecialCharTok{$}\NormalTok{rank}
\end{Highlighting}
\end{Shaded}

\begin{verbatim}
## [1] 3
\end{verbatim}

\begin{Shaded}
\begin{Highlighting}[]
\FunctionTok{qr}\NormalTok{(}\FunctionTok{cbind}\NormalTok{(}\FunctionTok{m}\NormalTok{(}\FloatTok{0.75}\NormalTok{), cb))}\SpecialCharTok{$}\NormalTok{rank}
\end{Highlighting}
\end{Shaded}

\begin{verbatim}
## [1] 3
\end{verbatim}

Next, we calculate \(2 D_3^{+} \left( B \otimes I_3 \right)\)

\begin{Shaded}
\begin{Highlighting}[]
\NormalTok{B }\OperatorTok{=}\NormalTok{ Matrix([[}\DecValTok{1}\NormalTok{, }\DecValTok{0}\NormalTok{], [}\OperatorTok{{-}}\NormalTok{k}\OperatorTok{*}\NormalTok{t, }\DecValTok{1}\NormalTok{], [}\OperatorTok{{-}}\NormalTok{t, }\DecValTok{0}\NormalTok{]])}

\CommentTok{\# Duplication matrix}
\NormalTok{D3 }\OperatorTok{=}\NormalTok{  Matrix([[}\DecValTok{1}\NormalTok{, }\DecValTok{0}\NormalTok{, }\DecValTok{0}\NormalTok{, }\DecValTok{0}\NormalTok{, }\DecValTok{0}\NormalTok{, }\DecValTok{0}\NormalTok{],}
\NormalTok{              [}\DecValTok{0}\NormalTok{, }\DecValTok{1}\NormalTok{, }\DecValTok{0}\NormalTok{, }\DecValTok{0}\NormalTok{, }\DecValTok{0}\NormalTok{, }\DecValTok{0}\NormalTok{],}
\NormalTok{              [}\DecValTok{0}\NormalTok{, }\DecValTok{0}\NormalTok{, }\DecValTok{1}\NormalTok{, }\DecValTok{0}\NormalTok{, }\DecValTok{0}\NormalTok{, }\DecValTok{0}\NormalTok{],}
\NormalTok{              [}\DecValTok{0}\NormalTok{, }\DecValTok{1}\NormalTok{, }\DecValTok{0}\NormalTok{, }\DecValTok{0}\NormalTok{, }\DecValTok{0}\NormalTok{, }\DecValTok{0}\NormalTok{],}
\NormalTok{              [}\DecValTok{0}\NormalTok{, }\DecValTok{0}\NormalTok{, }\DecValTok{0}\NormalTok{, }\DecValTok{1}\NormalTok{, }\DecValTok{0}\NormalTok{, }\DecValTok{0}\NormalTok{],}
\NormalTok{              [}\DecValTok{0}\NormalTok{, }\DecValTok{0}\NormalTok{, }\DecValTok{0}\NormalTok{, }\DecValTok{0}\NormalTok{, }\DecValTok{1}\NormalTok{, }\DecValTok{0}\NormalTok{],}
\NormalTok{              [}\DecValTok{0}\NormalTok{, }\DecValTok{0}\NormalTok{, }\DecValTok{1}\NormalTok{, }\DecValTok{0}\NormalTok{, }\DecValTok{0}\NormalTok{, }\DecValTok{0}\NormalTok{],}
\NormalTok{              [}\DecValTok{0}\NormalTok{, }\DecValTok{0}\NormalTok{, }\DecValTok{0}\NormalTok{, }\DecValTok{0}\NormalTok{, }\DecValTok{1}\NormalTok{, }\DecValTok{0}\NormalTok{],}
\NormalTok{              [}\DecValTok{0}\NormalTok{, }\DecValTok{0}\NormalTok{, }\DecValTok{0}\NormalTok{, }\DecValTok{0}\NormalTok{, }\DecValTok{0}\NormalTok{, }\DecValTok{1}\NormalTok{]])}

\CommentTok{\# Pseudo{-}Inverse of duplication matrix}
\NormalTok{D3p }\OperatorTok{=}\NormalTok{ (np.transpose(D3)}\OperatorTok{*}\NormalTok{D3).inv()}\OperatorTok{*}\NormalTok{np.transpose(D3)}

\CommentTok{\# Kronecker product of B with the 3{-}dimensional identity matrix}
\NormalTok{B\_otimes\_I3 }\OperatorTok{=}\NormalTok{ Matrix(np.kron(B, np.eye(}\DecValTok{3}\NormalTok{)))}


\DecValTok{2}\OperatorTok{*}\NormalTok{D3p}\OperatorTok{*}\NormalTok{B\_otimes\_I3}
\end{Highlighting}
\end{Shaded}

\begin{verbatim}
## [  2.0        0         0       0    0    0 ]
## [                                           ]
## [-1.0*k*t    1.0        0      1.0   0    0 ]
## [                                           ]
## [ -1.0*t      0        1.0      0    0    0 ]
## [                                           ]
## [   0      -2.0*k*t     0       0   2.0   0 ]
## [                                           ]
## [   0       -1.0*t   -1.0*k*t   0    0   1.0]
## [                                           ]
## [   0         0       -2.0*t    0    0    0 ]
\end{verbatim}

Last, we check that the matrix
\(\left(\begin{smallmatrix}2D_{n}^{+}(B\otimes I_{n})\\ C_{B} \end{smallmatrix}\right)\)
is of full column rank.

\begin{Shaded}
\begin{Highlighting}[]
\NormalTok{D3pBI3 }\OtherTok{\textless{}{-}} \ControlFlowTok{function}\NormalTok{(k, t) \{}

  \FunctionTok{matrix}\NormalTok{(}\FunctionTok{c}\NormalTok{(}\DecValTok{2}\NormalTok{, }\FunctionTok{rep}\NormalTok{(}\DecValTok{0}\NormalTok{, }\DecValTok{5}\NormalTok{), }\SpecialCharTok{{-}}\NormalTok{k}\SpecialCharTok{*}\NormalTok{t, }
           \DecValTok{1}\NormalTok{, }\DecValTok{0}\NormalTok{, }\DecValTok{1}\NormalTok{, }\DecValTok{0}\NormalTok{, }\DecValTok{0}\NormalTok{, }\SpecialCharTok{{-}}\NormalTok{t, }
           \DecValTok{0}\NormalTok{, }\DecValTok{1}\NormalTok{, }\FunctionTok{rep}\NormalTok{(}\DecValTok{0}\NormalTok{, }\DecValTok{3}\NormalTok{), }\DecValTok{0}\NormalTok{, }
           \SpecialCharTok{{-}}\DecValTok{2}\SpecialCharTok{*}\NormalTok{k}\SpecialCharTok{*}\NormalTok{t, }\DecValTok{0}\NormalTok{, }\DecValTok{0}\NormalTok{, }\DecValTok{2}\NormalTok{, }\DecValTok{0}\NormalTok{, }\DecValTok{0}\NormalTok{, }
           \SpecialCharTok{{-}}\NormalTok{t, }\SpecialCharTok{{-}}\NormalTok{k}\SpecialCharTok{*}\NormalTok{t, }\DecValTok{0}\NormalTok{, }\DecValTok{0}\NormalTok{, }\DecValTok{1}\NormalTok{, }
           \DecValTok{0}\NormalTok{, }\DecValTok{0}\NormalTok{, }\SpecialCharTok{{-}}\DecValTok{2}\SpecialCharTok{*}\NormalTok{t, }\FunctionTok{rep}\NormalTok{(}\DecValTok{0}\NormalTok{,}\DecValTok{3}\NormalTok{)),}\AttributeTok{byrow=}\ConstantTok{TRUE}\NormalTok{, }\AttributeTok{ncol=}\DecValTok{6}\NormalTok{)}
\NormalTok{\}}

\FunctionTok{qr}\NormalTok{(}\FunctionTok{rbind}\NormalTok{(}\FunctionTok{D3pBI3}\NormalTok{(}\AttributeTok{k=}\FloatTok{0.5}\NormalTok{, }\AttributeTok{t=}\FloatTok{0.75}\NormalTok{), CB))}\SpecialCharTok{$}\NormalTok{rank}
\end{Highlighting}
\end{Shaded}

\begin{verbatim}
## [1] 6
\end{verbatim}

\end{document}